\documentclass[11pt]{article}
\usepackage{amssymb,latexsym,amsmath,amsbsy}
\usepackage[dvips]{graphicx}
\usepackage{cite}
\newtheorem{theo}{Theorem}

\newtheorem{prop}[theo]{Proposition}
\def\deg{\mathop{\rm deg}\nolimits}
\def\ch{\mathop{\rm char}\nolimits}
\def\qdots{\mathinner{\mkern1mu\raise1pt\vbox{\kern7pt\hbox{.}}\mkern2mu \raise4pt\hbox{.}\mkern2mu\raise7pt\hbox{.}\mkern1mu}}

\begin{document}
\begin{center}
{\Large \bf
Gel'fand-Zetlin Basis and Clebsch-Gordan Coefficients for  Covariant  Representations  of the Lie superalgebra $\mathfrak{gl}(m|n)$ }\\[5mm] 
{\bf N.I.~Stoilova}\footnote{E-mail: Neli.Stoilova@UGent.be; Permanent address:
Institute for Nuclear Research and Nuclear Energy, Boul.\ Tsarigradsko Chaussee 72,
1784 Sofia, Bulgaria} {\bf and J.\ Van der Jeugt}\footnote{E-mail:
Joris.VanderJeugt@UGent.be}\\[1mm]
Department of Applied Mathematics and Computer Science,
Ghent University,\\
Krijgslaan 281-S9, B-9000 Gent, Belgium.  
\end{center}

\addtolength{\parskip}{2mm}

\begin{abstract}
A Gel'fand-Zetlin basis is introduced for the irreducible covariant tensor representations of
the Lie superalgebra $\mathfrak{gl}(m|n)$. 
Explicit expressions for the generators of the Lie superalgebra acting on this basis are determined.
Furthermore, Clebsch-Gordan coefficients corresponding to the tensor product of any covariant
tensor representation of $\mathfrak{gl}(m|n)$ with the natural representation $V([1,0,\ldots,0])$ of $\mathfrak{gl}(m|n)$ with
highest weight (1,0,\ldots,0) are computed. 
Both results are steps for the explicit construction of the parastatistics Fock space.
\end{abstract}


\setcounter{equation}{0}
\section{Introduction} \label{sec:Introduction}%
The representation theory of (basic) classical Lie (super)algebras plays a
central role in many branches of mathematics and physics. 
The first explicit constructions  of finite-dimensional irreducible representations
were given by Gel'fand and Zetlin~\cite{GZ, GZ-so}. 
They  introduced a basis in any finite-dimensional irreducible $\mathfrak{gl}(n)$ module $V$ 
considering  the chain of subalgebras $\mathfrak{gl}(n) \supset \mathfrak{gl}(n-1) \supset \ldots \supset \mathfrak{gl}(1) $.
Since each such module $V$ is a direct sum of irreducible $\mathfrak{gl}(n-1)$ modules $V=\sum_{i}\oplus V_i$, 
where the decomposition is multiplicity 
free, and any irreducible $\mathfrak{gl}(1)$  module $V(1)$ is a one dimensional space, 
the vectors corresponding to all possible flags 
$V\equiv V(n) \supset V(n-1) \supset \ldots \supset V(1)$ and labeled by the highest weights of 
$V(k)$, constitute a basis
in $V$. This basis is now called a Gel'fand-Zetlin (GZ) basis
in the $\mathfrak{gl}(n)$ module $V$~\cite{GZ}. 

In a similar way one can introduce a basis in each finite-dimensional $\mathfrak{so}(n)$ module~\cite{GZ-so} 
considering the chain of subalgebras
$
\mathfrak{so}(n) \supset \mathfrak{so}(n-1) \supset \ldots \supset \mathfrak{so}(2).
$ 
Contrary to $\mathfrak{gl}(n)$, where the basis consists of orthonormal weight vectors, the GZ-basis vectors for
$\mathfrak{so}(n)$~\cite{GZ-so} are not eigenvectors for the Cartan subalgebra
(so the GZ-basis vectors are not weight vectors).  

This approach does not work for the symplectic Lie algebras $\mathfrak{sp}(2n)$ since the restriction 
$\mathfrak{sp}(2n) \downarrow \mathfrak{sp}(2n-2)$ is not multiplicity free.
Since the papers of Gel'fand and Zetlin~\cite{GZ, GZ-so} were published in 1950, many different methods
were developed to construct bases in the modules of the classical Lie algebras (see for instance 
the review paper~\cite{Molev}). Finally, a complete solution of the problem for the $\mathfrak{sp}(2n)$ modules was  
given by Molev~\cite{Molev-sp} in 1999. He 
used  finite-dimensional irreducible representations of the so called twisted Yangians.  Molev applied his approach
also to the orthogonal Lie algebras~\cite{Molev-so-even, Molev-so-odd}. 
The new basis consists of weight vectors but in turn lacks the orthogonality property.
In such a way the problem to construct  a natural basis for the Lie algebras $\mathfrak{so}(n)$ and $\mathfrak{sp}(2n)$,
which accommodate both properties (weight vectors and orthogonality) remains an open one.  

Also some steps towards a generalization of
the concept of GZ-basis for basic classical Lie superalgebras have been taken (see~\cite{Palev, Palev2, palev89-2}). 
Irrespective of the progress, there is still
much to be done in order to complete the representation theory of the basic classical Lie superalgebras. 
In the present paper we take a step further in this respect introducing a Gel'fand-Zetlin basis in the irreducible 
covariant tensor representations of 
the general linear Lie superalgebra  $\mathfrak{gl}(m|n)$~\cite{Kac1,Kac2}
and writing down explicit expressions for the transformation of 
the basis vectors under the action of the algebra generators. In this case the Gel'fand-Zetlin  basis 
vectors accommodate both nice properties -- they are orthonormal and weight vectors.
Next, using the matrix elements of the $\mathfrak{gl}(m|n)$ covariant tensor representations, we compute certain 
Clebsch-Gordan coefficients of the Lie superalgebra $\mathfrak{gl}(m|n)$.

The motivation for the present work comes from some physical ideas. In 1953
Green~\cite{Green} introduced more general statistics than the common Fermi-Dirac and 
Bose-Einstein statistics, namely the parafermion and paraboson statistics. 
These generalizations have an algebraic formulation in terms of generators 
and relations.
The parafermion operators $f_j^\pm$,   satisfying
\begin{equation}
[[f_{ j}^{\xi}, f_{ k}^{\eta}], f_{l}^{\epsilon}]=\frac 1 2
(\epsilon -\eta)^2
\delta_{kl} f_{j}^{\xi} -\frac 1 2  (\epsilon -\xi)^2
\delta_{jl}f_{k}^{\eta},  
\label{f-rels}
\end{equation}
where $j,k,l\in \{1,2,\ldots,m\}$ and $\eta, \epsilon, \xi \in\{+,-\}$ (to be interpreted as $+1$ and $-1$
in the algebraic expressions $\epsilon -\xi$ and $\epsilon -\eta$), generate the Lie algebra 
$\mathfrak{so}(2m+1)$~\cite{Kamefuchi,Ryan}. 
Similarly, $n$ pairs of paraboson operators $b_j^\pm$, $j=1,2,\ldots, n$, satisfying
\begin{equation}
[\{ b_{ j}^{\xi}, b_{ k}^{\eta}\} , b_{l}^{\epsilon}]= (\epsilon -\xi) \delta_{jl} b_{k}^{\eta} 
 +  (\epsilon -\eta) \delta_{kl}b_{j}^{\xi}, 
\label{b-rels}
\end{equation}
generate  the orthosymplectic Lie superalgebra $\mathfrak{osp}(1|2n)$~\cite{Ganchev}. 
The paraboson and parafermion Fock spaces, characterized by a positive integer~$p$, often referred to 
as the order of statistics, are unitary lowest weight representations of the
relevant algebras with a nondegenerate lowest weight space (i.e.\ with a unique
vacuum). Despite  their importance, an explicit construction of the parafermion and paraboson
Fock spaces was not known until recently. 
For the case of parafermions, this explicit construction was 
given in~\cite{parafermion}, and for  parabosons in~\cite{paraboson}.

It is natural to extend these results to a system consisting of parafermions $f_j^\pm$ and
parabosons $b_j^\pm$. It was proved by Palev~\cite{Palev1} that the relative commutation relations between $m$ 
parafermions~(\ref{f-rels}) and $n$ parabosons~(\ref{b-rels}) can be defined in such a way that they generate the
orthosymplectic Lie superalgebra $\mathfrak{osp}(2m+1|2n)$. Then the parastatistics Fock space of order $p$ corresponds to 
an infinite-dimensional unitary representation of $\mathfrak{osp}(2m+1|2n)$ and it can be constructed explicitly 
using similar techniques as in~\cite{paraboson, parafermion}, namely using the branching $\mathfrak{osp}(2m+1|2n)\supset \mathfrak{gl}(m|n)$,
an induced representation construction,
a basis description for the covariant tensor representations of  $\mathfrak{gl}(m|n)$,
Clebsch-Gordan coefficients of $\mathfrak{gl}(m|n)$, and  the method of reduced matrix elements. Therefore in order to construct 
the parastatistics Fock space first we need the covariant tensor representations 
of the Lie superalgebra $\mathfrak{gl}(m|n)$ in an explicit form. Since it is easy to see that the triple relations~(\ref{f-rels})
and~(\ref{b-rels}) imply that the set $(f_1^+, \ldots, f_m^+, b_1^+, \ldots,b_n^+) $ is a standard $\mathfrak{gl}(m|n)$ 
tensor of rank $(1,0,\ldots,0)$, for the construction of the parastatistics Fock space one needs the $\mathfrak{gl}(m|n)$ 
Clebsch-Gordan coefficients corresponding to the tensor product $V([\mu]^{r})\otimes V([1,0,\ldots,0])$, where $V([\mu]^{r})$ is any 
$\mathfrak{gl}(m|n)$
irreducible covariant tensor representation and $V([1,0,\ldots,0])$ is the representation of 
$\mathfrak{gl}(m|n)$ with highest weight $(1,0,\ldots,0)$.
This paper deals with these two problems.  
In section~\ref{sec:glmn} we define the Lie superalgebra $\mathfrak{gl}(m|n)$ and remind the reader of some representation theory of 
$\mathfrak{gl}(m|n)$, in particular of the concept of typical, atypical and covariant tensor representations. In the next
section, we construct the covariant tensor representations of $\mathfrak{gl}(m|n)$ introducing a 
Gel'fand-Zetlin basis. We present the action of the $\mathfrak{gl}(m|n)$ generators on the  basis,
and give some indications of how we proved that the defining relations of the algebra are satisfied in these representations.
The computation of the Clebsch-Gordan coefficients in given in section~\ref{sec:CGCs}. 

\setcounter{equation}{0}
\section{The Lie superalgebra $\mathfrak{gl}(m|n)$} \label{sec:glmn}

The underlying vector space for the Lie superalgebra ${\mathfrak g}=\mathfrak{gl}(m|n)$ consists of the space of $(r\times r)$-matrices, with
\begin{equation}
r=m+n.
\end{equation}
The Lie superalgebra ${\mathfrak g}=\mathfrak{gl}(m|n)$ can be
defined~\cite{Kac1,Kac2} through its natural
matrix realization
\begin{equation}
\mathfrak{gl}(m|n)=\{ x=\left(\begin{array}{cc} A&B\\ C&D\end{array}\right)
| A\in M_{m\times m}, B\in M_{m\times n}, C\in M_{n\times m},
  D\in M_{n\times n} \},
\label{defgl}
\end{equation}
where $M_{p\times q}$ is the space of all $p\times q$ complex matrices.
The even subalgebra $\mathfrak{gl}(m|n)_{\bar 0}$ has $B=0$ and $C=0$; the odd
subspace $\mathfrak{gl}(m|n)_{\bar 1}$ has $A=0$ and $D=0$. 
Note that $\mathfrak{gl}(m|n)_{\bar 0} = \mathfrak{gl}(m)\oplus \mathfrak{gl}(n)$. We denote by $\mathfrak{gl}(m|n)_{+1}$ the space of matrices
$\left(\begin{array}{cc}0&B\\0&0\end{array}\right)$ and by
$\mathfrak{gl}(m|n)_{-1}$ the space of matrices
$\left(\begin{array}{cc}0&0\\ C&0\end{array}\right)$. Then
${\mathfrak g}=\mathfrak{gl}(m|n)$ has a ${\mathbb Z}$-grading which is consistent with the
${\mathbb Z}_2$-grading~\cite{scheunert79}, 
namely ${\mathfrak g}={\mathfrak g}_{-1}\oplus {\mathfrak g}_0\oplus
{\mathfrak g}_{+1}$ with ${\mathfrak g}_{\bar 0}={\mathfrak g}_0$ and ${\mathfrak g}_{\bar 1}={\mathfrak g}_{-1}\oplus {\mathfrak g}_{+1}$.
The Lie superalgebra is then defined by means of the bracket $[\![ x,y ]\!]=xy-(-1)^{\deg(x)\deg(y)}yx$, where $x$ and $y$ are homogeneous elements.

A basis for ${\mathfrak g}=\mathfrak{gl}(m|n)$ consists of matrices $e_{ij}$ ($i,j=1,2,\ldots,
r$) with entry $1$ at position $(i,j)$ and $0$ elsewhere.
Alternatively, the Lie superalgebra ${\mathfrak g}$ can be defined by means of generators and relations.
A Cartan subalgebra ${\mathfrak h}$ of ${\mathfrak g}$ is spanned by the elements $e_{jj}$
($j=1,2,\ldots,r$), and a set of generators of $\mathfrak{gl}(m|n)$ is given
by the Chevalley generators $h_j\equiv e_{jj}$ ($j=1,\ldots,r$), $e_i\equiv e_{i,i+1}$ and
$f_i\equiv e_{i+1,i}$ ($i=1,\ldots,r-1$). 
Then ${\mathfrak g}$ can be defined as the free associative
superalgebra over ${\mathbb C}$    and
generators $h_{j}$, ($j=1,2,\ldots,r$)
and $e_{i}$, $f_{i}$ ($i=1,2,\ldots,r-1$) subject to the following
relations~\cite{F,T,S} (unless stated otherwise, the indices below run over
all possible values):

\begin{itemize}
\item The Cartan-Kac relations:
\begin{align}
& [h_{i},h_{j}]=0;  \label{kk}
\\
& 
[h_{i},e_{j}]=(\delta_{ij}-\delta_{i,j+1})e_{j}; \label{hiej} \\
& [h_{i},f_{j}]=-(\delta_{ij}-\delta_{i,j+1})f_{j};
 \label{hifj} \\
& [e_{i}, f_{j}]=0 \ \hbox{ if } i\neq j; \label{eifj} \\
& [e_{i},f_{i}]=h_{i}-h_{i+1} 
 \ \hbox{ if } i\ne m; \label{eifi}\\
& \{e_{m},f_{m}\}=h_{m}+h_{m+1}; \label{enfn}
\end{align}

\item The Serre relations for the $e_{i}$:
\begin{align}
& e_{i}e_{j}=e_{j}e_{i} \hbox{ if } \vert i-j \vert \neq 1;\qquad
e_{m}^2=0; \label{ee}\\
&
e_{i}^2e_{i+1}-2e_{i}e_{i+1}e_{i}+e_{i+1}e_{i}^2=0,
\hbox{ for } i\in \{1,\ldots,m-1\}\cup\{m+1,\ldots,n+m-2\}; \label{eee1}\\
& e_{i+1}^2e_{i}-2e_{i+1}e_{i}e_{i+1}+e_{i}e_{i+1}^2=0,
\hbox{ for } i\in\{1,\ldots,m-2\}\cup\{m,\ldots,n+m-2\}; \label{eee2}\\
& e_{m}e_{m-1}e_{m}e_{m+1}+e_{m-1}e_{m}e_{m+1}e_{m}+e_{m}e_{m+1}e_{m}e_{m-1}\nonumber\\
 & +e_{m+1}e_{m}e_{m-1}e_{m}-2e_{m}e_{m-1}e_{m+1}e_{m}=0;
 \label{eeee}
 \end{align}

\item The relations obtained from (\ref{ee})--(\ref{eeee}) by replacing
every $e_{i}$ by $f_{i}$.
\end{itemize}

The space dual to ${\mathfrak h}$ is ${\mathfrak h}^*$ and
is described by the forms
$\epsilon_i$ ($i=1,\ldots,r$) where
$\epsilon_j:x\rightarrow A_{jj}$ for $1\leq j\leq m$ and $\epsilon_{m+j}:x\rightarrow
D_{jj}$ for $1\leq j\leq n$, and where $x$ is
given as in (\ref{defgl}). 
 The components of an
element $\Lambda\in {\mathfrak h}^*$ will be written as $[\mu]^r=[\mu_{1r},\mu_{2r},\ldots,\mu_{rr}]$
where $\Lambda=\sum_{i=1}^r \mu_{ir}\epsilon_i$ and $\mu_{ir}$ are complex numbers.
The elements of ${\mathfrak h}^*$ are called the weights. The roots of $\mathfrak{gl}(m|n)$
 take the
form $\epsilon_i-\epsilon_j$ ($i\ne j$); the positive roots are
those with $1\leq i<j\leq r$, and of importance are the $mn$ odd positive roots
\begin{equation}
\beta_{ip} = \epsilon_i-\epsilon_p, \quad\hbox{ with }1\leq i\leq m\hbox{ and }
 m+1\leq p \leq r.
\label{odd}
\end{equation}
$\Lambda\in {\mathfrak h}^*$ with components $[\mu]^r$ will be called an integral dominant
weight if $\mu_{ir}-\mu_{i+1,r}\in{\mathbb Z}_+=\{0,1,2,\ldots\}$ for all $i\ne m$
($1\leq i\leq r-1$). For every integral dominant weight $\Lambda\equiv[\mu]^r$
we denote by $V^0(\Lambda)$ the simple ${\mathfrak g}_0$ module with highest weight
$\Lambda$; this is simply the finite-dimensional
$\mathfrak{gl}(m)\oplus \mathfrak{gl}(n)$ module with $\mathfrak{gl}(m)$
labels $[\mu_{1r},\ldots \mu_{mr}]$ and with $\mathfrak{gl}(n)$ labels
$[\mu_{m+1,r},\ldots,\mu_{rr}]$. The module $V^0(\Lambda)$ can be extended
to a ${\mathfrak g}_0\oplus {\mathfrak g}_{+1}$ module by the requirement that ${\mathfrak g}_{+1}V^0(\Lambda)=0$.
The induced ${\mathfrak g}$ module $\overline V([\Lambda])$, first introduced by
Kac~\cite{Kac2} and usually referred to as the Kac-module, is defined
by
\begin{equation}
\overline V([\Lambda]) = \hbox{Ind}_{{\mathfrak g}_0\oplus {\mathfrak g}_{+1}}^{{\mathfrak g}} V^0(\Lambda)
 \cong U({\mathfrak g}_{-1})\otimes V^0(\Lambda),
\end{equation}
where $U({\mathfrak g}_{-1})$ is the universal enveloping algebra of ${\mathfrak g}_{-1}$.
It follows that $\dim \overline V([\Lambda]) = 2^{nm} \dim V^0(\Lambda)$.
By definition, $\overline V([\Lambda])$ is a highest weight module;
unfortunately, $\overline V([\Lambda])$ is not always a simple ${\mathfrak g}$ module.
It contains a unique maximal (proper) submodule $M[\Lambda]$, and the quotient
module
\begin{equation}
V([\Lambda])=\overline V([\Lambda])/M[\Lambda] \label{vl}
\end{equation}
is a finite-dimensional
simple module with highest weight $\Lambda$. In fact, Kac~\cite{Kac2}
proved the following:
\begin{theo} Every finite-dimensional simple ${\mathfrak g}$ module is isomorphic
to a module of type~(\ref{vl}), where $\Lambda\equiv[\mu]^r \equiv[\mu_{1r},\mu_{2,r},\ldots,\mu_{rr}]$ is integral dominant.
Moreover, every finite-dimensional simple ${\mathfrak g}$ module is uniquely
characterized by its integral dominant highest weight $\Lambda$.
\end{theo}
An integral dominant weight $\Lambda=[\mu]^r=[\mu_{1r},\mu_{2r},\ldots,\mu_{rr}]$ (resp.~$\overline V([\Lambda])$,
resp.~$V([\Lambda])$) is called a typical weight (resp.~a typical Kac
module, resp.~a typical simple module) if and only if $\langle
\Lambda+\rho|\beta_{ip}\rangle\ne 0$ for all odd positive roots
$\beta_{ip}$ of (\ref{odd}),
where $2\rho$ is the sum of all positive roots of ${\mathfrak g}$. Otherwise
$\Lambda$, $\overline V([\Lambda])$ and $V([\Lambda])$ are called atypical.
The importance of these definitions follows from another theorem of
Kac~\cite{Kac2}:
\begin{theo}
The Kac-module $\overline V([\Lambda])$ is a simple ${\mathfrak g}$ module if and only
if $\Lambda$ is typical.
\end{theo}
For an integral dominant highest weight $\Lambda=[\mu]^r$ it is convenient
to introduce the following labels~\cite{palev89-2}:
\begin{equation}
l_{ir}=\mu_{ir}-i+m+1,\quad(1\leq i \leq m);\qquad
l_{pr}=-\mu_{pr}+p-m,\quad(m+1\leq p\leq r).
\label{lir}
\end{equation}
In terms of these, one can deduce that $\langle\Lambda+\rho|\beta_{ip}\rangle=
l_{ir}-l_{pr}$, and hence the conditions for typicality take
a simple form.

Apart from the distinction between typical and atypical irreducible finite-dimensional modules of 
$\mathfrak{gl}(m|n)$, it is possible to distinguish between such modules on the basis of their relationship to
tensor modules of various kinds. For instance, Berele and Regev~\cite{BR},
showed that the tensor product $V([1,0,\ldots, 0])^{\otimes N}$ of $N$ copies of the natural $(m+n)$-dimensional representation
$V([1,0,\ldots, 0])$ of $\mathfrak{gl}(m|n)$ is completely reducible, and the irreducible components, $V_\lambda$,
can be labeled by a partition $\lambda$ of $N$ of length $l(\lambda)$ and weight $|\lambda|$, where 
$\lambda =(\lambda_1, \lambda_2, \ldots, \lambda_\ell)$, with $l(\lambda)=\ell$, $|\lambda|=\lambda_1+\lambda_2+\ldots+
\lambda_\ell=N$, and $\lambda_i\geq \lambda_{i+1}>0$ for $i=1,2,\ldots,\ell-1$, satisfying the condition $\lambda_{m+1}\leq n$.
For definitions regarding partitions, see~\cite{Macdonald}. 
The condition $\lambda_{m+1}\leq n$ is known as the {\em hook condition}: in terms of Young diagrams,
it means that the diagram of $\lambda$ should be inside the $(m,n)$-hook~\cite{BR}.
The representations thus obtained are  called  irreducible 
covariant tensor representations and  are necessarily finite dimensional. 
Then according to Theorem 1, there must exist an integral dominant weight $\Lambda^\lambda$ such that $V_{\lambda}$
is isomorphic to $V([\Lambda^\lambda])$. The relation between $\Lambda^\lambda$ $\equiv [\mu ]^r\equiv [\mu_{1r},
\mu_{2r},\ldots ,\mu_{rr}]$, $(r=m+n)$ and $\lambda =(\lambda_1, \lambda_2,\ldots)$ is such that~\cite{JHKR}: 
\begin{align}
& \mu_{ir}=\lambda_i, \quad 1\leq i\leq m, \label{hwpart1}\\
& \mu_{m+i,r}=\max\{0, \lambda'_i-m\}, \quad 1\leq i\leq n, \label{hwpart2}
\end{align}
where $\lambda'$ is the partition conjugate~\cite{Macdonald} to $\lambda$. 
Conversely if  $\Lambda$ $\equiv [\mu ]^r\equiv [\mu_{1r},\mu_{2r},\ldots ,\mu_{rr}]$ is integral dominant 
with all $\mu_{ir}\in{\mathbb Z}_+$ and 
\begin{equation}
\mu_{mr}\geq \# \{i:\mu_{ir}>0,\; m+1\leq i \leq r\},
\end{equation} 
then there exists a $\lambda$ such that $V([\Lambda])$ is isomorphic to the irreducible covariant tensor module
$V_\lambda$, and the components of $\lambda$ are given explicitly by
\begin{align}
& \lambda_i=\mu_{ir}, \quad 1\leq i\leq m, \label{hwpart3}\\
& \lambda_{m+i}=\#\{j: \mu_{m+j,r}\leq i, \;\; 1\leq j \leq n\}, \quad 1\leq i\leq n.\label{hwpart4}
\end{align}

The main feature of irreducible covariant tensor modules of $\mathfrak{gl}(m|n)$ is that their characters are known explicitly~\cite{BR, Sergeev}. 
Just as the characters of irreducible covariant tensor modules of $\mathfrak{gl}(m)$, which may be expressed in 
terms of ordinary Schur functions~\cite{Littlewood}, 
the characters of $\mathfrak{gl}(m|n)$ can be given in terms of supersymmetric Schur functions.
Following Macdonald~\cite{Macdonald}, the Schur function in the variables $({\mathbf x})=(x_1, x_2,\ldots, x_m)$ specified by the partition $\sigma$
is denoted by $s_\sigma({\mathbf x})$. Schur functions satisfy the following product and quotient rules:
\begin{align}
& s_\sigma({\mathbf x})s_\tau({\mathbf x})= \sum_{\lambda} c_{\sigma \tau}^\lambda s_{\lambda}({\mathbf x}) \\
& s_{\lambda/\tau}({\mathbf x})= \sum_{\sigma} c_{\sigma \tau}^\lambda s_{\sigma}({\mathbf x}),
\end{align}
where the coefficients $c_{\sigma \tau}^\lambda$ are the famous Littlewood-Richardson coefficients, 
and the summations are over partitions $\lambda$ and $\sigma$ with $|\lambda|= |\sigma|+|\tau|$. Berele and Regev~\cite{BR}
proved the following:
\begin{theo}
\label{T3}
Let $V([\Lambda^\lambda])$ be an irreducible $\mathfrak{gl}(m|n)$ covariant tensor module specified by a partition $\lambda$, and
let
\begin{align}
& x_i=e^{\epsilon_i}, \quad\quad  1\leq i \leq m,\nonumber\\
& y_i=e^{\epsilon_{m+i}}, \quad 1\leq i \leq n.\nonumber
\end{align}
Then the character of $V([\Lambda^\lambda])$ is given by
\[
\ch V([\Lambda^\lambda]) = s_\lambda({\mathbf x| \mathbf y}),
\]
where $s_\lambda({\mathbf x| \mathbf  y})$ is the supersymmetric Schur function of $({\mathbf x})=(x_1, x_2,\ldots, x_m)$ 
and $({\mathbf y})=(y_1, y_2,\ldots, y_n)$ defined by
\[
s_\lambda({\mathbf x| \mathbf y})= \sum_{\tau} s_{\lambda/\tau}({\mathbf x})s_{\tau'}({\mathbf y})
= \sum_{\sigma, \tau} c_{\sigma \tau}^\lambda s_{\sigma}({\mathbf x})s_{\tau'}({\mathbf y}),
\] 
with $l(\sigma)\leq m$ and $l(\tau')\leq n$.
\end{theo}

For the Lie algebra $\mathfrak{gl}(m)$, the simplicity of a Gel'fand-Zetlin basis stems from the fact that the decomposition
from $\mathfrak{gl}(m)$ to $\mathfrak{gl}(m-1)$ is so easy (and multiplicity free) for covariant tensor modules.
Since the characters of these $\mathfrak{gl}(m)$ modules are given by Schur functions $s_\lambda({\mathbf x})$,
this decomposition is deduced from the following formula~\cite{Macdonald}:
\begin{equation}
s_\lambda(x_1,\ldots,x_{m-1},x_m) = 
\sum_{\sigma} s_\sigma(x_1,\ldots,x_{m-1}) \cdot x_m^{|\lambda|-|\sigma|},
\label{s-reduction}
\end{equation}
where the sum is over all partitions $\sigma$ such that
\begin{equation}
\lambda_1 \geq \sigma_1 \geq \lambda_2 \geq \sigma_2 \geq \cdots
\geq \sigma_{m-1} \geq \lambda_m .
\label{ineq-s}
\end{equation}
These last inequalities give rise to the so-called in-betweenness conditions in $\mathfrak{gl}(m)$ GZ-patterns.
In terms of the notions introduced in~\cite{Macdonald}, \eqref{ineq-s} means that $\lambda-\sigma$ is
a {\em horizontal strip}.

Various interesting expressions also exist for supersymmetric Schur functions~\cite{King}.
In particular, there is also a combinatorial expression in terms of supersymmetric tableaux of shape~$\lambda$.
{}From this expression (or from the one in Theorem~\ref{T3}), one deduces the following result:
\begin{equation}
s_\lambda(x_1,\ldots,x_m|y_1,\ldots,y_{n-1},y_n) = 
\sum_{\sigma} s_\sigma(x_1,\ldots,x_m|y_1,\ldots,y_{n-1}) \cdot y_n^{|\lambda|-|\sigma|},
\label{ss-reduction}
\end{equation}
where the sum is now over all partitions $\sigma$ in the $(m,n-1)$-hook such that
\begin{equation}
\lambda_1' \geq \sigma_1' \geq \lambda_2' \geq \sigma_2' \geq \cdots
\geq \sigma_{\ell-1}' \geq \lambda_\ell'\;,
\label{ineq-ss}
\end{equation}
where $\ell=\lambda_1$ is the length of $\lambda'$.
In terms of the notions of~\cite{Macdonald}, $\lambda-\sigma$ is a {\em vertical strip}.
This expression will be particularly useful when decomposing the covariant tensor representation of $\mathfrak{gl}(m|n)$ 
characterized by $\lambda$ in terms of $\mathfrak{gl}(m|n-1)$ representations.

\setcounter{equation}{0}
\section{Covariant tensor representations of  $\mathfrak{gl}(m|n)$} \label{sec:representations} 

Let
$V([\mu]^{r})$ be an irreducible covariant tensor module of $\mathfrak{gl}(m|n)$, namely the nonnegative 
integer $r$-tuple   
\begin{equation}
[\mu]^{r}=[\mu_{1r}, \mu_{2r}, \ldots , \mu_{rr}],  \label{mr}
\end{equation}
is such that
\begin{equation}
\mu_{ir}-\mu_{i+1,r}\in {\mathbb Z}_+, \; \forall i\neq m, \; i =1,\ldots, r-1 \label{cond1}
\end{equation}
and 
\begin{equation}
\mu_{mr}\geq \# \{i:\mu_{ir}>0,\; m+1\leq i \leq r\}. \label{cond2}
\end{equation}
Within a given $\mathfrak{gl}(m|n)$ module $V([\mu]^{r})$ the numbers~(\ref{mr}) are fixed.

For covariant tensor representations of the Lie algebra $\mathfrak{gl}(m)$, the relation between the partition
characterizing the highest weight and the highest weight itself is straightforward.
Moreover, the decomposition from $\mathfrak{gl}(m)$ to $\mathfrak{gl}(m-1)$ for such representations is very easy,
following~\eqref{s-reduction}. That is why the GZ-basis vectors for $\mathfrak{gl}(m)$ have such a simple pattern.

For covariant tensor representations of the Lie superalgebra $\mathfrak{gl}(m|n)$ the situation is different.
First of all, the relation between the partition $\lambda$ characterizing the highest weight and
the components of the highest weight is more involved, see~\eqref{hwpart1}-\eqref{hwpart2}.
Therefore the conditions on the highest weight components, \eqref{cond1}-\eqref{cond2} are more complicated.
Still, it is necessary to use highest weight components in the labeling of basis vectors, in order to
describe the action of generators appropriately.
Secondly, the decomposition from $\mathfrak{gl}(m|n)$ to $\mathfrak{gl}(m|n-1)$ for covariant tensor representations
is fairly easy to describe using the partition labeling, according to~\eqref{ss-reduction}.
However, we need to translate this to the corresponding highest weights. 
This gives rise to the following propositions.
\begin{prop}
Consider the $\mathfrak{gl}(m|n)$ module $V([\mu]^{r})$ as a $\mathfrak{gl}(m|n-1)$ module.
Then $V([\mu]^{r})$ can be represented as a direct sum of covariant simple $\mathfrak{gl}(m|n-1)$ modules,
\begin{equation}
V([\mu]^{r})=\sum_i \oplus V_i([\mu]^{r-1}), \label{gl(mn-1)}
\end{equation}
where 
\begin{itemize}
\item[I.] All $V_i([\mu]^{r-1})$ carry inequivalent representations of $\mathfrak{gl}(m|n-1)$
\begin{align}
& [\mu]^{r-1}=[\mu_{1,r-1}, \mu_{2,r-1},\ldots , \mu_{r-1,r-1}], \\
& \mu_{i,r-1}-\mu_{i+1,r-1}\in{\mathbb Z}_+,\;\forall i\neq m, \; i =1,\ldots, r-2,\\
&  \mu_{m,r-1}\geq \# \{i:\mu_{i,r-1}>0,\; m+1\leq i \leq r-1\} .
\end{align}
\item[II.] 
\begin{equation}
 \begin{array}{rl}
1.& \mu_{ir}-\mu_{i,r-1}=\theta_{i,r-1}\in\{0,1\},\quad 1\leq i\leq m,\\
2.& \mu_{i,r}-\mu_{i,r-1}\in{\mathbb Z}_+\hbox{ and }\; \mu_{i,r-1}-\mu_{i+1,r}\in{\mathbb Z}_+,\quad
    m+1\leq i\leq r-1.
 \end{array}
\label{cond0}
\end{equation}
\end{itemize}
\end{prop}

\begin{prop}
Consider a covariant $\mathfrak{gl}(m|1)$ module $V([\mu]^{m+1})$ as a $\mathfrak{gl}(m)$ module.
Then $V([\mu]^{m+1})$ can be represented as a direct sum of simple $\mathfrak{gl}(m)$ modules,
\begin{equation}
V([\mu]^{m+1})=\sum_i \oplus V_i([\mu]^m), \label{gl(m)}
\end{equation}
where 
\begin{itemize}
\item[I.] All $V_i([\mu]^m)$ carry inequivalent representations of $\mathfrak{gl}(m)$
\begin{equation}
[\mu]^{m}=[\mu_{1m}, \mu_{2m},\ldots , \mu_{mm}], \; \mu_{im}-\mu_{i+1,m}\in{\mathbb Z}_+  .\label{m}
\end{equation}
\item[II.] 
\begin{equation}
 \begin{array}{rl}
1.& \mu_{i,m+1}-\mu_{im}=\theta_{im}\in\{0,1\},\quad 1\leq i\leq m,\\
2.& \hbox{if }\; 
\mu_{m,m+1}=0, \hbox{ then}\; \theta_{mm}=0.
 \end{array}
\label{cond}
\end{equation}
\end{itemize}
\end{prop}

Using Proposition~1, Proposition~2 and the $\mathfrak{gl}(m)$ GZ-basis we have:
 
\begin{prop}
The set of vectors 
\begin{equation}
|\mu)\equiv |\mu)^r = \left|
\begin{array}{lclllcll}
\mu_{1r} & \cdots & \mu_{m-1,r} & \mu_{mr} & \mu_{m+1,r} & \cdots & \mu_{r-1,r}
& \mu_{rr}\\
\mu_{1,r-1} & \cdots & \mu_{m-1,r-1} & \mu_{m,r-1} & \mu_{m+1,r-1} & \cdots
& \mu_{r-1,r-1} & \\
\vdots & \vdots &\vdots &\vdots & \vdots & \qdots & & \\
\mu_{1,m+1} & \cdots & \mu_{m-1,m+1} & \mu_{m,m+1} & \mu_{m+1,m+1} & & & \\
\mu_{1m} & \cdots & \mu_{m-1,m} & \mu_{mm} & & & & \\
\mu_{1,m-1} & \cdots & \mu_{m-1,m-1} & & & & & \\
\vdots & \qdots & & & & & & \\
\mu_{11} & & & & & & &
\end{array}
\right)
= \left| \begin{array}{l} [\mu]^r \\[2mm] |\mu)^{r-1} \end{array} \right)
\label{mn}
\end{equation}
satisfying the conditions
\begin{equation}
 \begin{array}{rl}
1. & \mu_{ir}\in{\mathbb Z}_+ \; \hbox{are fixed and }  \mu_{jr}-\mu_{j+1,r}\in{\mathbb Z}_+ , \;j\neq m,\;
    1\leq j\leq r-1,\\
    & \mu_{mr}\geq \# \{i:\mu_{ir}>0,\; m+1\leq i \leq r\};\\
2.& \mu_{ip}-\mu_{i,p-1}\equiv\theta_{i,p-1}\in\{0,1\},\quad 1\leq i\leq m;\;
    m+1\leq p\leq r;\\
3.  &   \mu_{mp}\geq \# \{i:\mu_{ip}>0,\; m+1\leq i \leq p\}, \quad  m+1\leq p\leq r ;\\   
4.& \hbox{if  }\;
\mu_{m,m+1}=0, \hbox{then}\; \theta_{mm}=0;  \\
5.& \mu_{ip}-\mu_{i+1,p}\in{\mathbb Z}_+,\quad 1\leq i\leq m-1;\;
    m+1\leq p\leq r-1;\\
6.& \mu_{i,j+1}-\mu_{ij}\in{\mathbb Z}_+\hbox{ and }\mu_{i,j}-\mu_{i+1,j+1}\in{\mathbb Z}_+,\\
  &  1\leq i\leq j\leq m-1\hbox{ or } m+1\leq i\leq j\leq r-1.
 \end{array}
\label{cond3}
\end{equation}
constitute a basis in $V([\mu]^{r})$.
\end{prop}

The last condition corresponds to the in-betweenness condition and
ensures that the triangular pattern to the right of the $n\times m$
rectangle $\mu_{ip}$ ($1\leq i\leq m$; $m+1\leq p\leq r$) in (\ref{mn})
corresponds to a classical GZ-pattern for $\mathfrak{gl}(n)$,
and that the triangular pattern below this rectangle corresponds
to a GZ-pattern for $\mathfrak{gl}(m)$.

We shall refer to the basis~(\ref{mn}) as the GZ-basis for the covariant $\mathfrak{gl}(m|n)$ representations.
The task is now to give the explicit action of the $\mathfrak{gl}(m|n)$ Chevalley generators 
on the basis vectors~(\ref{mn}). Let  $|\mu)_{\pm ij}$ be the pattern obtained
from $|\mu)$ by replacing the entry $\mu_{ij}$ by $\mu_{ij}\pm 1$, and for the notations 
$l_{ij}$ see formula~(\ref{lir}).

The following is one of the two main results of this paper:

\begin{theo} The transformation of the irreducible covariant tensor
module $V([\mu]^{r})$ under the action of the $\mathfrak{gl}(m|n)$ generators is given by:
\begin{align}
& h_{k}|\mu)=\left(\sum_{j=1}^k \mu_{jk}-\sum_{j=1}^{k-1} \mu_{j,k-1}\right)|\mu), 
\quad 1\leq k\leq r; \label{e_kk}\\
& e_{k}|\mu)=\sum_{j=1}^k \left(-\frac{
{\prod_{i=1}^{k+1} (l_{i,k+1}-l_{jk})
\prod_{i=1}^{k-1} (l_{i,k-1}-l_{jk}-1)}}
{\prod_{i\neq j=1}^k (l_{ik}-l_{jk})
(l_{ik}-l_{jk}-1) } \right)^{1/2}|\mu)_{jk},\; \nonumber\\
& \hskip 6cm 1\leq k\leq m-1;\label{ek}\\
& f_{k}|\mu)=\sum_{j=1}^k \left(-
\frac{\prod_{i=1}^{k+1} (  l_{i,k+1}-l_{jk}+1  )
\prod_{i=1}^{k-1} (  l_{i,k-1}-l_{jk}  )}
{\prod_{i\neq j=1}^k (  l_{ik}-l_{jk}+1  )
(  l_{ik}-l_{jk}  ) } \right)^{1/2}|\mu)_{-jk},\nonumber\\
&\hskip 6cm 1\leq k\leq m-1;\label{fk} \\
& e_{m}|\mu)=\sum_{i=1}^m \theta_{im}(-1)^{i-1}
(-1)^{\theta_{1m}+ \ldots +\theta_{i-1,m} }(l_{i,m+1}-l_{m+1,m+1})^{1/2}\nonumber\\
& \times \left(\frac {\prod_{k=1}^{m-1} (  l_{k,m-1}-l_{i,m+1} )}
{\prod_{k\neq i=1}^m (  l_{k,m+1}-l_{i,m+1} )}
\right)^{1/2} |\mu)_{im};\label{em}\displaybreak\\
& f_{m}|\mu)=\sum_{i=1}^m (1-\theta_{im})(-1)^{i-1}
(-1)^{\theta_{1m}+ \ldots +\theta_{i-1,m} }(l_{i,m+1}-l_{m+1,m+1})^{1/2}\nonumber\\
 &\times\left(\frac{\prod_{k=1}^{m-1} (  l_{k,m-1}-l_{i,m+1} )}
{\prod_{k\neq i=1}^m (  l_{k,m+1}-l_{i,m+1} )}
\right)^{1/2} |\mu)_{-im};\label{fm}\\
& e_{p}|\mu)=\sum_{i=1}^m \theta_{ip} (-1)^{\theta_{1p}+ \ldots
+\theta_{i-1,p}+\theta_{i+1,p-1}+ \ldots +\theta_{m,p-1} }
(1-\theta_{i,p-1})\nonumber\\
&\times\prod_{k\neq i=1}^m \left(
\frac{(l_{i,p+1}-l_{kp})  (l_{i,p+1}-l_{kp}-1)} {
(l_{i,p+1}-l_{k,p+1})(l_{i,p+1}-l_{k,p-1}-1)} \right)^{1/2}\nonumber\\
&\times \left(\frac{\prod_{q=m+1}^{p-1}(l_{i,p+1}-l_{q,p-1}-1)
\prod_{q=m+1}^{p+1}(l_{i,p+1}-l_{q,p+1})}{ \prod_{q=m+1}^p
 (l_{i,p+1}-l_{qp}-1)(l_{i,p+1}-l_{qp})}\right)^{1/2} |\mu)_{ip} \nonumber\\
 &+\sum_{s=m+1}^p \left( - \frac{\prod_{q=m+1}^{p-1}(l_{q,p-1}-l_{sp}+1)
 \prod_{q=m+1}^{p+1}(l_{q,p+1}-l_{sp})} { \prod_{q\ne s=m+1}^p
 (l_{qp}-l_{sp})(l_{qp}-l_{sp}+1) } \right)^{1/2} \label{ep}\\
&\times \prod_{k=1}^m \left(\frac{(l_{kp}-l_{sp}) (l_{kp}-l_{sp}+1)} {
(l_{k,p+1}-l_{sp})(l_{k,p-1}-l_{sp}+1)}\right)^{1/2} |\mu)_{sp},
\quad m+1\leq p \leq r-1;\nonumber \\
& f_{p}|\mu)=\sum_{i=1}^m \theta_{i,p-1} (-1)^{\theta_{1p}+ \ldots
+\theta_{i-1,p}+\theta_{i+1,p-1}+ \ldots +\theta_{m,p-1} }(1-\theta_{ip})\nonumber\\
&\times\prod_{k\neq i=1}^m \left(\frac{
(l_{i,p+1}-l_{kp})  (l_{i,p+1}-l_{kp}-1)}{
(l_{i,p+1}-l_{k,p+1})(l_{i,p+1}-l_{k,p-1}-1)} \right)^{1/2}
 \nonumber\\
 &\times \left(\frac{\prod_{q=m+1}^{p-1}(l_{i,p+1}-l_{q,p-1}-1)
\prod_{q=m+1}^{p+1}(l_{i,p+1}-l_{q,p+1})}{ \prod_{q=m+1}^p
 (l_{i,p+1}-l_{qp}-1)(l_{i,p+1}-l_{qp})}\right)^{1/2} |\mu)_{-ip} \nonumber\\
 &+
 \sum_{s=m+1}^p \left( -\frac
 {\prod_{q=m+1}^{p-1} (l_{q,p-1}-l_{sp}) \prod_{q=m+1}^{p+1}
 (l_{q,p+1}-l_{sp}-1)}{ \prod_{q\ne s=m+1}^p
(l_{qp}-l_{sp}-1)(l_{qp}-l_{sp}) }\right)^{1/2}  \label{fp}\\
&\times \prod_{k=1}^m \left(\frac{(l_{kp}-l_{sp}-1) (l_{kp}-l_{sp})} {
(l_{k,p+1}-l_{sp}-1)(l_{k,p-1}-l_{sp})}\right)^{1/2} |\mu)_{-sp},
\quad m+1\leq p \leq r-1;\nonumber 
\end{align}
\end{theo}
In the above expressions, $\sum_{k\ne i=1}^m$ or
$\prod_{k\ne i=1}^m$ means that
$k$ takes all values from $1$ to $m$ with $k\ne i$. If a vector
from the right hand side of (\ref{e_kk})-(\ref{fp}) does not belong to the module
under consideration, then the corresponding term is zero even if the
coefficient in front is undefined; if an equal number of factors in
numerator and denominator are simultaneously equal to zero, they
should be canceled out. 

To conclude this section, we shall make some comments on the proof of this theorem.
In order to prove that the explicit actions~(\ref{e_kk})-(\ref{fp}) give a representation of 
$\mathfrak{gl}(m|n)$ it is sufficient to show that~(\ref{e_kk})-(\ref{fp}) satisfy the relations~(\ref{kk})-(\ref{eeee})
 (plus the $f$-Serre relations).
The irreducibility then follows from the fact that for any nonzero vector $x\in V([\mu]^{r})$ there exists
a polynomial ${\cal P}$ of $\mathfrak{gl}(m|n)$ generators such that ${\cal P}$$x=V([\mu]^{r})$.

To show that~(\ref{kk})-(\ref{eifj}) are satisfied is  straightforward. The difficult Cartan-Kac relations 
to be verified are~(\ref{eifi}) and~(\ref{enfn}). For instance relation~(\ref{enfn}), with the actions~(\ref{e_kk})-(\ref{fp}),
is valid if and only if
\begin{equation}
\sum_{i=1}^m (l_{i,m+1}-l_{m+1,m+1}) \frac{\prod_{k=1}^{m-1}
(l_{k,m-1}-l_{i,m+1})} { \prod_{k\ne i=1}^m (l_{k,m+1}-l_{i,m+1})}
= \sum_{k=1}^{m-1}(l_{k,m+1}-l_{k,m-1})+l_{m,m+1}
-l_{m+1,m+1} .
\end{equation}
The proof of this relation is given in~\cite{PSJ}.
For the $e$- and $f$-Serre relations, the calculations are lengthy,
but  collecting terms with the same Gel'fand-Zetlin basis
vector and  taking apart the common factors, the remaining
factor always reduces to a very simple algebraic expression like:
\begin{equation}
a(b+1)-(a+1)b=a-b,\quad
\frac{1}{a(a-1)}+\frac{1}{a(a+1)}=\frac{2}{(a-1)(a+1)},
\end{equation}
from which the validity follows.

\setcounter{equation}{0}
\section{Clebsch-Gordan coefficients  of  $\mathfrak{gl}(m|n)$} \label{sec:CGCs}

In this section we compute the Clebsch-Gordan coefficients  of  $\mathfrak{gl}(m|n)$ corresponding
to the tensor product $V([\mu]^r)\otimes V([1,0,\ldots,0])$ of any irreducible $\mathfrak{gl}(m|n)$ 
covariant tensor module $V([\mu]^r)$ with the natural $(m+n)$-dimensional $\mathfrak{gl}(m|n)$  representation $V([1,0,\ldots,0])$.
It is well known and it is easy to see from the character formula that:
\begin{equation}
V([\mu]^r)\otimes V([1,0,\ldots,0]) =\sum_{k=1}^r \oplus V([\mu]_{+k}^r), \label{tensprod}
\end{equation}
where $[\mu]_{+k}^r$ is obtained from $[\mu]^r$ by the replacement of $\mu_{kr}$ by 
$\mu_{kr}+1$ and on the right hand side of~(\ref{tensprod}) the summands for which the 
conditions~(\ref{cond1})-(\ref{cond2}) are not fulfilled are omitted. 
We choose two orthonormal bases in the space~(\ref{tensprod}):
\begin{align}
& \left| \begin{array}{l} [\mu]^r \\[2mm] |\mu)^{r-1} \end{array} \right)
\otimes |1_j)\in V([\mu]^r)\otimes V([1,0,\ldots,0]) \quad {\rm and} \\
&  \nonumber \\
&\left| \begin{array}{l} [\mu]^r_{+k} \\[2mm] |\mu^\prime)^{r-1} \end{array} \right)\in V([\mu]_{+k}^r), \;\; k=1,\ldots,r,
\end{align}
where the vectors $\left| \begin{array}{l} [\mu]^r \\[2mm] |\mu)^{r-1} \end{array} \right)$ and 
$\left| \begin{array}{l} [\mu]^r_{+k} \\[2mm] |\mu^\prime)^{r-1} \end{array} \right)$ satisfy the 
conditions of Proposition 6, and 
$|1_j), \; j=1,\ldots, r$ is a pattern which 
consists of $r-j$ zero rows at the bottom (denoted by $0\cdots0=\dot 0$), and the first $j$ rows are of the form $1 0 \cdots 0$
(denoted by $1\dot 0$).
Then in general
\begin{equation}
\left| \begin{array}{l} [\mu]^r_{+k} \\[2mm] |\mu^\prime)^{r-1} \end{array} \right)=
\sum_{|\mu )^{r}, |1_j)} 
  \left( \begin{array}{ll} [\mu]^r \\[2mm] |\mu)^{r-1} \end{array} ; \right.
 \begin{array}{l}1 0 \cdots 0 0\\[-1mm]
1 0 \cdots 0\\[-1mm]  \cdots\\[-1mm] 0 \end{array}  
\left| \begin{array}{ll} [\mu]^r_{+k} \\[2mm] |\mu^\prime)^{r-1} \end{array} \right)
\left| \begin{array}{l} [\mu]^r \\[2mm] |\mu)^{r-1} \end{array} \right)
 \otimes |1_j),\label{CGCs}
\end{equation}
where $\left( \begin{array}{ll} [\mu]^r \\[2mm] |\mu)^{r-1} \end{array} ; \right.
 \begin{array}{l}1 0 \cdots 0 0\\[-1mm]
1 0 \cdots 0\\[-1mm]  \cdots\\[-1mm] 0 \end{array}\left| \begin{array}{ll} [\mu]^r_{+k} \\[2mm] |\mu^\prime)^{r-1} \end{array} \right) 
\equiv 
\left( \begin{array}{ll} [\mu]^r \\[2mm] |\mu)^{r-1} \end{array} ; \right.
 \begin{array}{l}\\[-1mm]
|1_j)\\[-1mm] \\ \end{array}\left| \begin{array}{ll} [\mu]^r_{+k} \\[2mm] |\mu^\prime)^{r-1} \end{array} \right)
$ are the Clebsch-Gordan coefficients (CGCs).
Acting onto both sides of relation~(\ref{CGCs}) by the Cartan generators $h_i, \;i=1,\ldots,r$
and taking into account formula~(\ref{e_kk}) it follows that the CGC of $\mathfrak{gl}(m|n)$ vanishes if one
of the relations 
\begin{align}
& \sum_{s=1}^{p} \mu_{sp}^\prime =\sum_{s=1}^{p} \mu_{sp}, \quad p=1,\ldots , r-j,\\
& \sum_{s=1}^{p} \mu_{sp}^{\prime}= \sum_{s=1}^{p} \mu_{sp}+1, \quad p=r+1-j, \ldots , r-1
\end{align}
is not fulfilled.

Since multiple representations are absent in~(\ref{tensprod}) we have  for the CGCs 
\begin{align}
& \left( \begin{array}{ll} [\mu]^r \\[2mm] |\mu)^{r-1} \end{array} ; \right.
 \begin{array}{l}1 0 \cdots 0 0\\[-1mm]
1 0 \cdots 0\\[-1mm]  \cdots\\[-1mm] 0 \end{array}\left| \begin{array}{ll} [\mu]^r_{+k} \\[2mm] |\mu^\prime)^{r-1} \end{array} \right)
 \nonumber\\
& = 
\left( \begin{array}{l}  [\mu]^r \\ {[\mu]}^{r-1} \end{array} \right.
\left| \begin{array}{l} 1\dot{0} \\ \epsilon \dot{0} \end{array} \right|
\left. \begin{array}{l} [\mu]^r_{+k}  \\ {[\mu^\prime]}^{r-1} \end{array} \right) \times
\left( \begin{array}{ll} [\mu]^{r-1} \\[2mm] |\mu)^{r-2} \end{array} ; \right.
 \begin{array}{l}1 0 \cdots 0 0\\[-1mm]
1 0 \cdots 0\\[-1mm]  \cdots\\[-1mm] 0 \end{array}\left| \begin{array}{ll} [\mu^{\prime}]^{r-1} \\[2mm] |\mu^\prime)^{r-2} \end{array} \right) \label{IsoCGC}. 
\end{align}
In the right hand side, the first factor is an isoscalar factor~\cite{Vilenkin}, and the second factor
is a CGC of $\mathfrak{gl}(m|n-1)$.
The middle pattern in the $\mathfrak{gl}(m|n-1)$ CGC is that of the $\mathfrak{gl}(m|n)$ CGC with the first row deleted.
The middle pattern in the isoscalar factor consists of the first two rows of the middle pattern in the
left hand side, so $\epsilon$ is 0 or 1. 
If $\epsilon=0$, then $[\mu^\prime]^{r-1}=[\mu]^{r-1}$.
If $\epsilon=1$ then $[\mu^\prime]^{r-1}=[\mu_{1,r-1},\ldots, \mu_{s,r-1}+1,\ldots,\mu_{r-1,r-1}]=[\mu]^{r-1}_{+s}$
for some $s$-value.

In addition to all this we followed the general procedure for computing Clebsch-Gordan coefficients.
First, the highest weight vector of the irreducible module $V([\mu]_{+1}^r)$ is equal 
to the tensor product of the two highest weight vectors of the components of the tensor product 
$V([\mu]^r)\otimes V([1,0,\ldots,0])$. Then any other vector in the same irreducible module $V([\mu]_{+1}^r)$
is obtained 
by acting with polynomials of negative root vectors on this vector. The highest weight vector in the 
irreducible module $V([\mu]_{+2}^r)$ is  (up to a phase) fixed by the requirement that
it is orthogonal to the unique vector in $V([\mu]_{+1}^r)$ with the same weight, namely 
$[\mu_{1r},\mu_{2r}+1,\mu_{3r},\mu_{4r},\ldots,\mu_{rr}]$ as of the highest 
weight vector in this second space $V([\mu]_{+2}^r)$. Then again,
all vectors in the irreducible module  $V([\mu]_{+2}^r)$ are found by the actions of polynomials 
of negative root vectors of the algebra to the corresponding highest weight vector of $V([\mu]_{+2}^r)$.
Next the highest weight vector of $V([\mu]_{+3}^r)$ has to be orthogonal to all vectors in $V([\mu]_{+1}^r)$
and $V([\mu]_{+2}^r)$ with weight $(\mu_{1r},\mu_{2r},\mu_{3r}+1,\mu_{4r},\ldots,\mu_{rr})$ (the highest weight of $V([\mu]_{+3}^r)$).
Note that following this general procedure for computing CGCs one should have in mind  two other important
facts from  representation theory of Lie superalgebras.
First: Lie superalgebra representation spaces are   ${\mathbb Z}_2$-graded spaces and  for the considered irreducible
$\mathfrak{gl}(m|n)$ covariant tensor modules $V([\mu]^r)=V_{\bar{0}}([\mu]^r)\oplus V_{\bar{1}}([\mu]^r)$ there are 
two possibilities for the ${\mathbb Z}_2$-grading, namely
$|\mu )\in V_{\bar{0}}([\mu]^r) $ (resp.\ $|\mu )\in V_{\bar{1}}([\mu]^r) $) if 
$\sum_{i=1}^m \sum_{p=m+1}^r\theta_{i,p-1}=\sum_{i=1}^m \sum_{p=m+1}^r(\mu_{i,p}-\mu_{i,p-1})$ is even (resp.\ odd).
The first grading will be referred to as the natural grading, and the other one as the opposite grading.
Second: 
The action of a Lie superalgebra   generator $g$ in a tensor product of two ${\mathfrak g}$-modules $V$ and $W$ is given by
$$
g(x\otimes y)= gx\otimes y + (-1)^{\deg (g) \deg (x)} x \otimes gy, \;\; x\in V, \;\; y\in W,
$$
and in the computations only the grading of the first module $V$ plays role. Because of this we fix that 
in~(\ref{tensprod}) $V([1,0,\ldots,0])$ has the natural grading.
As a consequence, the vectors $|1_j)$ of $V([1,0,\ldots,0])$ have the following degree:
\begin{equation}
\deg |1_j) = \bar 1 \;\hbox{ if } \; 1\leq j \leq n, \qquad
\deg |1_j) = \bar 0 \; \hbox{ if } \; n+1\leq j \leq n+m.
\end{equation}
The degree of the vectors is important since in general the vector 
$|\mu )^{r-1}$ from the $\mathfrak{gl}(m|n-1)$ module does not necessarily have the same grading as the 
vector $|\mu )^{r}$ from the $\mathfrak{gl}(m|n)$ module.

Iterating the  procedure for computing Clebsch-Gordan coefficients and the corresponding isoscalar factors,
it is clear that there are two distinct cases. 
First, when $1\leq j \leq n$, one will finally reach a trivial CGC of $\mathfrak{gl}(m|n-j)$ which is equal to~$1$ 
because the middle pattern consist of zeros only;
in this case the computed $\mathfrak{gl}(m|n)$ CGC is a product of isoscalar factors only.
Secondly, when $n+1\leq j \leq n+m$, the iteration leads to a product of isoscalar factors times a CGC of $\mathfrak{gl}(m)$.
For these simple $\mathfrak{gl}(m)$ CGCs, there exist closed form expressions, see e.g.~\cite{Vilenkin,paraboson}.
Thus we reach to the following:
\begin{theo}
The Clebsch-Gordan coefficients corresponding
to the tensor product 
\[
V([\mu]^r)\otimes V([1,0,\ldots,0])
\] 
of an  irreducible $\mathfrak{gl}(m|n)$ 
covariant tensor module $V([\mu]^r)$ with the natural $(m+n)$-dimensional $\mathfrak{gl}(m|n)$  representation $V([1,0,\ldots,0])$
are

\begin{itemize}
\item products of isoscalar factors (for $j=1,\ldots,n$) 

\begin{align}
& \left( \begin{array}{ll} [\mu]^r \\[2mm] |\mu)^{r-1} \end{array} ; \right.
 \begin{array}{l}\\[-1mm]
|1_j)\\[-1mm] \\ \end{array}\left| \begin{array}{ll} [\mu]^r_{+k} \\[2mm] |\mu^\prime)^{r-1} \end{array} \right)
 = \xi (-1)^{\sum_{q=1}^j\sum_{i=1}^m\theta_{i,r-q}}
\left( \begin{array}{l}  [\mu]^r \\ {[\mu]}^{r-1} \end{array} \right.
\left| \begin{array}{l} 1\dot{0} \\ 1 \dot{0} \end{array} \right|
\left. \begin{array}{l} [\mu]^r_{+k}  \\ {[\mu^\prime]}^{r-1} \end{array} \right)\times \ldots \nonumber\\
& \times \left( \begin{array}{l}  [\mu]^{r+2-j} \\ {[\mu]}^{r+1-j} \end{array} \right.
\left| \begin{array}{l} 1\dot{0} \\ 1 \dot{0} \end{array} \right|
\left. \begin{array}{l} [\mu\prime]^{r+2-j}  \\ {[\mu^\prime]}^{r+1-j} \end{array} \right) \left( \begin{array}{l}  [\mu]^{r+1-j} \\ {[\mu]}^{r-j} \end{array} \right.
\left| \begin{array}{l} 1\dot{0} \\ 0 \dot{0} \end{array} \right|
\left. \begin{array}{l} [\mu\prime]^{r+1-j}  \\ {[\mu^\prime]}^{r-j} \end{array} \right)\times 1  
 \label{CGC1},
 \end{align} 
where $\xi =(-1)^{\deg([\mu]^r)}$,  $\deg([\mu]^r)$ being the degree of the highest weight vector of $V([\mu]^r)$;
  
\item products of isoscalar factors and a $\mathfrak{gl}(m)$ CGC~\cite{Vilenkin,paraboson} (for $j=n+1,\ldots,r$)
\begin{align}
& \left( \begin{array}{ll} [\mu]^r \\[2mm] |\mu)^{r-1} \end{array} ; \right.
 \begin{array}{l}\\[-1mm]
|1_j)\\[-1mm] \\ \end{array}\left| \begin{array}{ll} [\mu]^r_{+k} \\[2mm] |\mu^\prime)^{r-1} \end{array} \right)
  = 
\left( \begin{array}{l}  [\mu]^r \\ {[\mu]}^{r-1} \end{array} \right.
\left| \begin{array}{l} 1\dot{0} \\ 1 \dot{0} \end{array} \right|
\left. \begin{array}{l} [\mu]^r_{+k}  \\ {[\mu^\prime]}^{r-1} \end{array} \right) \times  \ldots 
\nonumber\\
&\times \left( \begin{array}{l}  [\mu]^{m+1} \\ {[\mu]}^{m} \end{array} \right.
\left| \begin{array}{l} 1\dot{0} \\ 1 \dot{0} \end{array} \right|
\left. \begin{array}{l} [\mu\prime]^{m+1}  \\ {[\mu^\prime]}^{m} \end{array} \right)
\left( \begin{array}{ll} [\mu]^m \\[2mm] |\mu)^{m-1} \end{array} ; \right.
 \begin{array}{l}\\[-1mm]
|1_{j-n})\\[-1mm] \\ \end{array}\left| \begin{array}{ll} [\mu\prime ]^m \\[2mm] |\mu^\prime)^{m-1} \end{array} \right)
 \label{CGC2};  
\end{align}
\end{itemize}
and the isoscalar factors are given by:
\begin{align}
&\left( \begin{array}{l} [\mu]^{r} \\ {[\mu]}^{r-1} \end{array} \right.
\left| \begin{array}{l} 1 \dot{0} \\0 \dot{0} \end{array} \right|
\left. \begin{array}{l} [\mu]^{r}_{+k} \\ {[\mu]}^{r-1} \end{array} \right)
\nonumber\\
&
= (-1)^{k-1}(-1)^{\sum_{i=k}^m}\theta_{i,r-1}
\left( \prod_{i\neq k=1}^m \left(\frac{l_{kr}-l_{ir}+1}{l_{kr}-l_{i,r-1}}
\right)  
 \frac{\prod_{p=m+1}^{r-1}  
(l_{kr}-l_{p,r-1} )}
{ \prod_{p=m+1}^{r} (l_{kr}-l_{pr}+1)}
\right)^{1/2}
1\leq k \leq m;\label{liso1} \displaybreak\\
&\nonumber\\
&\left( \begin{array}{l} [\mu]^{r} \\ {[\mu]}^{r-1} \end{array} \right.
\left| \begin{array}{l} 1 \dot{0} \\0 \dot{0} \end{array} \right|
\left. \begin{array}{l} [\mu]^{r}_{+k} \\ {[\mu]}^{r-1} \end{array} \right)
= \left( \prod_{i=1}^m \left(\frac{l_{ir}-l_{kr}}{l_{i,r-1}-l_{kr}+1}
\right)  
 \frac{\prod_{p=m+1}^{r-1}  
(l_{p,r-1}-l_{kr}+1 )}
{ \prod_{p\neq k=m+1}^{r} (l_{pr}-l_{kr})}
\right)^{1/2}\nonumber\\
& \hskip 8cm  \quad m+1\leq k \leq r;\label{liso2}\\
&\nonumber\\
&\left( \begin{array}{l} [\mu]^{r} \\ {[\mu]}^{r-1} \end{array} \right.
\left| \begin{array}{l} 1 \dot{0} \\1 \dot{0} \end{array} \right|
\left. \begin{array}{l} [\mu]^{r}_{+k} \\ {[\mu]}^{r-1}_{+q} \end{array} \right)
=(-1)^{k+q}(-1)^{\sum_{i=\min(k+1,q+1)}^{\max(k-1,q-1)}}\theta_{i,r-1}S(k,q) \nonumber\\
&\times \left( \prod_{i\neq k,q=1}^m 
\frac{(l_{i,r-1}-l_{k,r-1}-1-\delta_{kq}+2\theta_{i,r-1})
(l_{i,r-1}-l_{q,r-1})}{(l_{ir}-l_{kr})(l_{ir}-l_{qr})}
\right)^{\frac{\theta_{q,r-1}}{2}} \nonumber\\  
&
\times\frac{1}{(l_{kr}-l_{qr})^{1-\delta_{kq}}}
\left( \prod_{p=m+1}^{r} \left(
\frac{l_{qr}-l_{pr}}
{l_{kr}-l_{pr}+1}
\right)
\prod_{p=m+1}^{r-1} \left(
\frac{l_{kr}-l_{q,r-1}}
{l_{q,r-1}-l_{p,r-1}}
\right)
\right)^{\frac{\theta_{q,r-1}}{2}}
 1\leq k,q \leq m;\label{liso3}\\
&\nonumber\\
&\left( \begin{array}{l} [\mu]^{r} \\ {[\mu]}^{r-1} \end{array} \right.
\left| \begin{array}{l} 1 \dot{0} \\1 \dot{0} \end{array} \right|
\left. \begin{array}{l} [\mu]^{r}_{+k} \\ {[\mu]}^{r-1}_{+q} \end{array} \right)
=(-1)^{k}(-1)^{\sum_{i=1}^{k-1}}\theta_{i,r-1}
\left(\frac{1}{l_{kr}-l_{q,r-1}}\right)^{1/2}
\nonumber\\
&\times \left( \prod_{i\neq k=1}^m \left(\frac{(l_{i,r-1}-l_{k,r-1}-1+2\theta_{i,r-1})
(l_{i,r-1}-l_{q,r-1}+1)}{(l_{ir}-l_{kr})(l_{ir}-l_{q,r-1})}
\right)\right)^{1/2} \nonumber\\  
&
\times
\left( \prod_{p=m+1}^{r} \left(\frac{|l_{pr}-l_{q,r-1}|}
{(l_{kr}-l_{pr}+1)}
\right) \prod_{p\neq q=m+1}^{r-1} \left(\frac{l_{kr}-l_{p,r-1}}
{|l_{p,r-1}-l_{q,r-1}+1|}
\right)\right)^{1/2}\nonumber\\
& \hskip 9cm 1\leq k \leq m, \quad m+1\leq q \leq r-1;\label{liso4}\\
&\nonumber\\
&\left( \begin{array}{l} [\mu]^{r} \\ {[\mu]}^{r-1} \end{array} \right.
\left| \begin{array}{l} 1 \dot{0} \\1 \dot{0} \end{array} \right|
\left. \begin{array}{l} [\mu]^{r}_{+k} \\ {[\mu]}^{r-1}_{+q} \end{array} \right)
=(-1)^{q}(-1)^{\sum_{i=q+1}^{m}}\theta_{i,r-1}
\left(\frac{1}{l_{qr}-l_{kr}+1}\right)^{1/2}
\nonumber\\
&\times \left( \prod_{i=1}^m \left(\frac{l_{ir}-l_{kr}}{l_{i,r-1}-l_{kr}+1}
\right) \prod_{i\neq q=1}^m \Big|\frac{l_{q,r-1}-l_{i,r-1}}{l_{qr}-l_{ir}}
 \Big|
 \prod_{p\neq k=m+1}^{r} \Big|\frac{l_{qr}-l_{pr}}{l_{pr}-l_{kr}}
\Big|
 \prod_{p=m+1}^{r-1} \Big|\frac{l_{p,r-1}-l_{kr}+1}{l_{qr}-l_{p,r-1}-1}
 \Big|\right)^{1/2}\nonumber\\
 & \hskip 9cm m+1\leq k \leq r, \quad 1\leq q \leq m;\label{liso5}\\
&\nonumber\\ 
&\left( \begin{array}{l} [\mu]^{r} \\ {[\mu]}^{r-1} \end{array} \right.
\left| \begin{array}{l} 1 \dot{0} \\1 \dot{0} \end{array} \right|
\left. \begin{array}{l} [\mu]^{r}_{+k} \\ {[\mu]}^{r-1}_{+q} \end{array} \right)
=S(k,q) \left( \prod_{i=1}^m \left(\frac{(l_{ir}-l_{kr})
(l_{i,r-1}-l_{q,r-1}+1)}{(l_{i,r-1}-l_{kr}+1)(l_{ir}-l_{q,r-1})}
\right)\right)^{1/2} \nonumber\\ 
&
\times
\left( \prod_{p\neq k=m+1}^{r} \Big|\frac{l_{pr}-l_{q,r-1}}
{l_{pr}-l_{kr}}\Big|
 \prod_{p\neq q=m+1}^{r-1} \Big| \frac{l_{p,r-1}-l_{kr}+1}
{l_{p,r-1}-l_{q,r-1}+1}
\Big| \right)^{1/2} 
m+1\leq k \leq r, \quad m+1\leq q \leq r-1;\label{liso6}\\
&\nonumber\\
& S(k,q) = \left\{ \begin{array}{rcl}
 {1} & \hbox{for} & k\leq q  \\ 
 {-1} & \hbox{for} & k>q.
 \end{array}\right.
 \label{S}
\end{align} 
\label{T8}
\end{theo}

The expressions in this Theorem look fairly complicated, however they are easy to use in practice. 
Let us give an example, and apply Theorem~\ref{T8} to the Lie superalgebra $\mathfrak{gl}(2|3)$, both for the case $j\leq n$ and $j>n$.
\begin{align}
& \left( \begin{array} {lllll} \mu_{15} \;\;\;\quad \;  \mu_{25}\;\;\;\quad \;  \mu_{35}\; \mu_{45}\; \mu_{55}   \\ 
\mu_{15}-1 \;\mu_{25}-1 \;\; \mu_{34}\; \mu_{44} \\
\mu_{15}-2 \;\mu_{25}-1 \;\; \mu_{33} \\
\mu_{15}-3 \;\mu_{25}-1 \\ \mu_{11}         \end{array} ; \right.
 \begin{array}{l}1 0 0 0 0\\[-1mm]
1 0 0 0\\[-1mm] 000 \\[-1mm] 00 \\ 0 \end{array}\left| 
\begin{array}{ll} 
\mu_{15} \;\;\;\quad \;  \mu_{25}+1\;\;  \mu_{35}\; \mu_{45}\; \mu_{55}   \\ 
\mu_{15}-1 \;\;\mu_{25} \;\;\;\quad \; \mu_{34}\; \mu_{44} \\
\mu_{15}-2 \;\;\mu_{25}-1 \;\; \mu_{33} \\
\mu_{15}-3 \;\;\mu_{25}-1 \\ \mu_{11}  
\end{array} \right)
 \nonumber\\
 &\nonumber\\
& = \xi
\left( \begin{array}{lllll} \mu_{15} \;\;\;\quad \;  \mu_{25}\;\;\;\quad \;  \mu_{35}\; \mu_{45}\; \mu_{55}   \\ 
\mu_{15}-1 \;\mu_{25}-1 \;\; \mu_{34}\; \mu_{44} \end{array} \right.
\left| \begin{array}{l} 1\dot{0} \\ 1 \dot{0} \end{array} \right|
\left. \begin{array}{l} 
\mu_{15} \;\;\;\quad \;  \mu_{25}+1\;\;  \mu_{35}\; \mu_{45}\; \mu_{55}   \\ 
\mu_{15}-1 \;\; \mu_{25} \;\;\;\quad \;  \mu_{34}\; \mu_{44}
\end{array} \right) 
\nonumber\\
&\nonumber\\
&\times (-1)^{\theta_{14}+\theta_{24}}
\left( \begin{array}{lllll} \mu_{15}-1 \;\;  \mu_{25}-1\;\;  \mu_{34}\; \mu_{44}   \\ \mu_{15}-2 \;\;\mu_{25}-1 \;\; \mu_{33} \end{array} \right.
\left| \begin{array}{l} 1\dot{0} \\  \dot{0} \end{array} \right|
\left. \begin{array}{l} 
\mu_{15}-1 \;\;  \mu_{25}\;\;\;\quad \;   \mu_{34}\; \mu_{44}   \\ \mu_{15}-2 \;\; \mu_{25}-1 \;\;  \mu_{33}
\end{array} \right)
\times 1 \nonumber\\
&\nonumber\\
&=-\xi \sqrt{     
\frac{(\mu_{25}+\mu_{35})(\mu_{25}+\mu_{45}-1)(\mu_{25}+\mu_{55}-2)(\mu_{25}+\mu_{33}-1)}
{(\mu_{25}+\mu_{35}+1)(\mu_{25}+\mu_{45})(\mu_{25}+\mu_{55}-1)(\mu_{25}+\mu_{34}-1)(\mu_{25}+\mu_{44}-2)}
}
\label{IsoCGC2}. 
\end{align}

\begin{align}
& \left( \begin{array} {lllll} \mu_{15} \;\;\;\quad \;  \mu_{25}\;\;\;\quad \;  \mu_{35}\; \mu_{45}\; \mu_{55}   \\ 
\mu_{15}-1 \;\mu_{25}-1 \;\; \mu_{34}\; \mu_{44} \\
\mu_{15}-2 \;\mu_{25}-1 \;\; \mu_{33} \\
\mu_{15}-3 \;\mu_{25}-1 \\ \mu_{11}         \end{array} ; \right.
 \begin{array}{l}1 0 0 0 0\\[-1mm]
1 0 0 0\\[-1mm] 100 \\[-1mm] 10 \\ 0 \end{array}\left| 
\begin{array}{ll} 
\mu_{15} \;\;\;\quad \;  \mu_{25}+1\;\;  \mu_{35}\; \mu_{45}\; \mu_{55}   \\ 
\mu_{15}-1 \;\;\mu_{25} \;\;\;\quad \; \mu_{34}\; \mu_{44} \\
\mu_{15}-2 \;\;\mu_{25} \;\;\;\quad \; \mu_{33} \\
\mu_{15}-3 \;\;\mu_{25} \\ \mu_{11}  
\end{array} \right)
 \nonumber\\
 &\nonumber\\
& = 
\left( \begin{array}{lllll} \mu_{15} \;\;\;\quad \;  \mu_{25}\;\;\;\quad \;  \mu_{35}\; \mu_{45}\; \mu_{55}   \\ 
\mu_{15}-1 \;\mu_{25}-1 \;\; \mu_{34}\; \mu_{44} \end{array} \right.
\left| \begin{array}{l} 1\dot{0} \\ 1 \dot{0} \end{array} \right|
\left. \begin{array}{l} 
\mu_{15} \;\;\;\quad \;  \mu_{25}+1\;\;  \mu_{35}\; \mu_{45}\; \mu_{55}   \\ 
\mu_{15}-1 \;\; \mu_{25} \;\;\;\quad \;  \mu_{34}\; \mu_{44}
\end{array} \right) 
\nonumber\\
&\nonumber\\
&\times 
\left( \begin{array}{lllll} \mu_{15}-1 \;\;  \mu_{25}-1\;\;  \mu_{34}\; \mu_{44}   \\ \mu_{15}-2 \;\;\mu_{25}-1 \;\; \mu_{33} \end{array} \right.
\left| \begin{array}{l} 1\dot{0} \\ 1 \dot{0} \end{array} \right|
\left. \begin{array}{l} 
\mu_{15}-1 \;\;  \mu_{25}\;\;   \mu_{34}\; \mu_{44}   \\ \mu_{15}-2 \;\; \mu_{25} \;\;  \mu_{33}
\end{array} \right)
 \nonumber\\
&\nonumber\\ 
&\times 
\left( \begin{array}{lllll} \mu_{15}-2 \;\;  \mu_{25}-1\;\;  \mu_{33}   \\ \mu_{15}-3 \;\;\mu_{25}-1  \end{array} \right.
\left| \begin{array}{l} 1\dot{0} \\ 1 \dot{0} \end{array} \right|
\left. \begin{array}{l} 
\mu_{15}-2 \;\;  \mu_{25}\;\;   \mu_{33}   \\ \mu_{15}-3 \;\; \mu_{25}
\end{array} \right) 
\nonumber\\
&\nonumber\\ 
&\times 
\left( \begin{array}{lllll} \mu_{15}-3 \;\;  \mu_{25}-1   \\ \mu_{11}  \end{array} ; \right.
\left. \begin{array}{l} 10 \\ 0 \end{array} \right|
\left. \begin{array}{l} 
\mu_{15}-3 \;\;  \mu_{25}   \\ \mu_{11}
\end{array} \right) \nonumber\\
&= \sqrt{     
\frac{(\mu_{25}+\mu_{35})(\mu_{25}+\mu_{45}-1)(\mu_{25}+\mu_{55}-2)(\mu_{25}+\mu_{34})(\mu_{25}+\mu_{44}-1)(\mu_{11}-\mu_{25}+1)}
{(\mu_{25}+\mu_{35}+1)(\mu_{25}+\mu_{45})(\mu_{25}+\mu_{55}-1)(\mu_{25}+\mu_{34}-1)(\mu_{25}+\mu_{44}-2)(\mu_{15}-\mu_{25}-1)}
}
\label{IsoCGC3}. 
\end{align}

\section*{Acknowledgments}
N.I.\ Stoilova was supported by  project P6/02 of the Interuniversity Attraction Poles Programme (Belgian State -- 
Belgian Science Policy).


\begin{thebibliography}{99}

\bibitem{GZ}
I.M. Gel'fand, M.L. Zetlin, 
 ``Finite-Dimensional Representations of the Group of Unimodular Matrices,''
Dokl.\ Akad.\ Nauk SSSR {\bf 71}, 825-828 (1950).
 
  
\bibitem{GZ-so}
I.M. Gel'fand, M.L. Zetlin, 
``Finite-dimensional representations of groups of orthogonal matrices,''
Dokl.\ Akad.\ Nauk SSSR  {\bf 71}, 1017-1020 (1950).

\bibitem{Molev} 
A.I. Molev, ``Gel'fand-Tsetlin bases for classical Lie algebras,''
Handbook of Algebra  {\bf 4}, 109-170 (2006). 
 
\bibitem{Molev-sp} 
A.I. Molev, ``A basis for representations of symplectic Lie algebras,''
Commun. Math. Phys.  {\bf 201}, 591-618 (1999).

\bibitem{Molev-so-even} 
A.I. Molev, ``A weight basis for representations of even orthogonal  Lie algebras,''
in \emph{Combinatorial Methods in Representation Theory}, Adv. Studies in Pure Math.   {\bf 28}, 223-242 (2000). 

\bibitem{Molev-so-odd} 
A.I. Molev, ``Weight bases of Gel'fand-Tsetlin type for representations of classical  Lie algebras,''
J. Phys.A: Math. Gen. {\bf 33}, 4143-4168 (2000).

\bibitem{Palev}
T.D. Palev, ``Irreducible finite-dimensional representations of the Lie superalgebra $gl(1,n)$ in a Gel'fand-Zetlin basis,''
Funct. Anal. Appl.  {\bf 21}, 245-246 (1987).

\bibitem{Palev2}
T.D. Palev, ``Irreducible finite-dimensional representations of the Lie superalgebra $gl(1,n)$ in a Gel'fand-Zetlin basis,''
J. Math. Phys.   {\bf 30},  1433-1442 (1989).

\bibitem{palev89-2}
 T.D. Palev, ``Essentially typical representations of the Lie 
 superalgebra $gl(n/m)$ in a Gel'fand-Zetlin basis,''
 Funkt.\ Anal.\ Prilozh.~{\bf 23}, No.~2,
        69-70 (1989) (in Russian);
 Funct.\ Anal.\ Appl.~{\bf 23}, 141-142 (1989) (English translation).



\bibitem{Kac1}
V.G. Kac, ``Lie superalgebras,''  Adv.\ Math. {\bf 26}, 8-96 (1977).

\bibitem{Kac2}
V.G. Kac,  ``Representations of classical Lie superalgebras,'' Lect.\ Notes Math. {\bf 676}, 597 (1978). 

\bibitem{Green}
H.S. Green,  
``A Generalized Method of Field Quantization,''
Phys.\ Rev. {\bf 90}, 270-273 (1953).

\bibitem{Kamefuchi}
S. Kamefuchi, Y. Takahashi, 
``A generalization of field quantization and statistics,''
Nucl.\ Phys. {\bf 36}, 177-206 (1962).

\bibitem{Ryan}
C. Ryan, E.C.G. Sudarshan, 
``Representations of parafermi rings,''
Nucl.\ Phys.\ {\bf 47}, 207-211 (1963).
 
\bibitem{Ganchev}
A.Ch. Ganchev,  T.D. Palev,   
``A Lie Superalgebraic Interpretation of the Para-Bose Statistics,''
J.\ Math.\ Phys. {\bf 21}, 797-799 (1980).
 
\bibitem{parafermion}
N.I. Stoilova, J. Van der Jeugt,   
``The parafermion Fock space and explicit $\mathfrak{so}(2n+1)$ representations,''
J. Phys A: Math. Theor. {\bf 41}, 075202 (13 pp) (2008).

\bibitem{paraboson}
S. Lievens, N.I. Stoilova, J. Van der Jeugt,  
``The paraboson Fock space and unitary irreducible representations of the Lie superalgebra $\mathfrak{osp}(1|2n)$,''
Commun.\ Math.\ Phys.\ {\bf  281}, 805-826 (2008).

\bibitem{Palev1}
T.D. Palev,  
``Para-Bose and para-Fermi operators as generators of orthosymplectic Lie superalgebras,''
J.\ Math.\ Phys. {\bf 23}, 1100-1102 (1982).
 
\bibitem{scheunert79}
M. Scheunert, ``The theory of Lie superalgebras,'' Lect. Notes in Mathematics, Vol. {\bf 716} (Berlin,
Heidelberg, New York: Springer, 1979).  
 
  
\bibitem{F}
R. Floreanini, D.A. Leites, L. Vinet, ``On the defining relations
 of quantum superalgebras,''
 Lett.\ Math.\ Phys.~{\bf 23}, 127-131 (1991).  
 
\bibitem{T} 
S.M. Khoroshkin, V.N. Tolstoy, ``Universal R-matrix for quantized 
 (super)algebras,''
 Commun.\ Math.\ Phys.~{\bf 141}, 599-617 (1991).

\bibitem{S}
M. Scheunert, ``Serre-type relations for special linear Lie superalgebras,''
 Lett.\ Math.\ Phys.~{\bf 24}, 173-181 (1992).
 



\bibitem{BR}
R. Berele, A. Regev, ``Hook Young diagrams, combinatorics and representations of Lie superalgebras,'' Bull. Amer. Math. Soc. (N.S.) 
{\bf 8}, No.~2,  337-339 (1983); R. Berele, A. Regev, ``Hook Young diagrams with applications to combinatorics and to representations of Lie superalgebras,'' 
Adv. in Math. {\bf 64}, No.~2, 118-175 (1987). 

\bibitem{Macdonald}
I.G. Macdonald, \emph{Symmetric Functions and Hall Polynomials} (Oxford U.P., Oxford, 1979).


\bibitem{JHKR}
J. Van der Jeugt, J.W.B. Hughes, R.C. King,   and  J. Thierry-Mieg,  
``Character formulas for irreducible modules of the Lie superalgebras $sl(m/n)$,''  
J.\ Math. Phys. {\bf 18}, 2278-2304 (1990).  


\bibitem{Sergeev}
A.N. Sergeev, ``The tensor algebra of the identity representation as a module over the Lie superalgebra $gl(n,m)$
 and $Q(n)$,''  Math. USSR Sbornik, {\bf 51}, 419-427 (1985).

 

\bibitem{Littlewood}
D.E. Littlewood, \emph{The theory of group characters} (Oxford U.P., Oxford, 1940). 


\bibitem{King}
R.C. King, ``$S$-functions and characters of Lie algebras and superalgebras,''
 \emph{Invariant theory and tableaux} (Minneapolis, MN, 1988),  226-261, IMA Vol. Math. Appl., 19  (Springer, New York, 1990).

\bibitem{PSJ}
T.D. Palev, N.I. Stoilova,  and J. Van der Jeugt,  
``Finite-Dimensional Representations of the Quantum Superalgebra  $U_q[gl(m/n)]$
and Related $q$-Identities,''   
Commun.\ Math.\ Phys.\ {\bf  166}, 367-378 (1994).
 
\bibitem{Vilenkin}
N.J. Vilenkin, A.U. Klimyk, \emph{Representation of Lie Groups and Special Functions,} Vol. 3:
\emph{Classical and Quantum Groups and Special Functions},  
(Kluwer Academic Publishers 1992).

 
\end{thebibliography}
\end{document}